# Probing Excitons in Ultrathin PbS Nanoplatelets with Enhanced Near-Infrared Emission

*Francisco Manteiga Vázquez,[†] Qianli Yu,[†] Lars F. Klepzig,[+,||] Laurens D. A. Siebbeles,[†] Ryan W. Crisp,[†,‡] Jannika Lauth [+,||,†,*]*

[†]Chemical Engineering Department, Delft University of Technology, Van der Maasweg 9, NL-2629 HZ Delft, The Netherlands

[+]Institute of Physical Chemistry and Electrochemistry, Leibniz Universität Hannover, Callinstr. 3A, D-30167 Hannover, Germany

[||]Cluster of Excellence PhoenixD (Photonics, Optics, and Engineering – Innovation Across Disciplines), Hannover, Germany

[‡]Chemistry of Thin Film Materials, Department of Chemistry and Pharmacy, Friedrich-Alexander University Erlangen-Nürnberg, Cauerstr. 3, D-91058, Erlangen, Germany

Corresponding Author

*Dr. Jannika Lauth, jannika.lauth@pci.uni-hannover.de

ABSTRACT

Strongly quantum-confined 2D colloidal PbS nanoplatelets (NPLs) are highly interesting materials for near-infrared optoelectronic applications. Here, we use ultrafast transient optical absorption spectroscopy to study the characteristics and dynamics of photoexcited excitons in ultrathin PbS NPLs with a cubic (rock-salt) structure. The NPLs are synthesized at near room temperature from lead oleate and thiourea precursors and show an optical absorption onset at 680 nm (1.8 eV) as well as photoluminescence at 720 nm (1.7 eV). By treating PbS NPLs with $CdCl_2$ in a post-synthetic step, their photoluminescence quantum yield is strongly enhanced from 1.4 % to 19.4 %. The surface treatment leads to an increased lead to sulfur ratio in the structures and associated reduced non-radiative recombination. Exciton-phonon interactions in pristine and $CdCl_2$ treated PbS NPLs at frequencies of 1.8 and 2.2 THz are apparent from coherent oscillations in the measured transient absorption spectra. This study is an important step forward in unraveling and controlling the optical properties of IV-VI semiconductor NPLs.

**TOC**

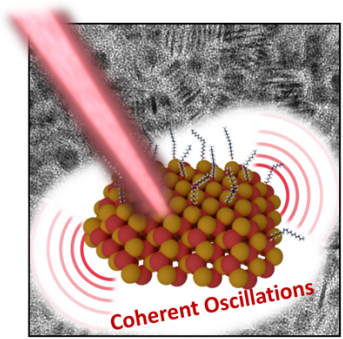

PbS nanoplatelets (NPLs), ultrafast transient optical absorption spectroscopy, near-infrared photoluminescence

Ultrathin 2D colloidal semiconductor nanosheets (NSs) and nanoplatelets (NPLs) are highly topical materials with interesting optoelectronic properties strongly different from their spherical nanocrystal (NC) or solid-state counterparts. In contrast to II-VI semiconductor NPLs (e.g. CdSe) exhibiting strong absorption and photoluminescence (PL) in the visible range, IV-VI semiconductors are of interest due to their widely tunable band gap from the far-infrared to the near-infrared (NIR) by variation of the thickness of the NSs. In addition, exciton binding energies and charge carrier mobilities can also be tuned by the thickness of the NSs. For example, thin PbS NSs exhibit stable excitons with high binding energies (~70 meV for 4 nm thickness) and carrier multiplication (CM) efficiencies approaching 100 %, meaning that all photons with energy exceeding the CM threshold generate additional electron-hole pairs.[1-3] Charge carrier mobilities increase with the thickness of the PbS NSs and become as high as 1000 cm$^2$/Vs for 16 nm thick NSs, which is close to the bulk mobility of PbS, while absorbing more light at longer wavelengths than thin NSs. Typically, PbS NSs with lateral dimensions of several $\mu$m and a thickness in the range of 4 – 40 nm are obtained by synthesis methods based on oriented attachment mechanisms.[4-6] Recently, new colloidal synthesis methods for further decreasing the thickness of 2D PbS layers down to atomically thin PbS NPLs have been implemented, including a combination of surface ligand stabilization[7] as well as kinetically controlled reaction conditions.[8-9] These methods yield ultrathin 2D PbS layers with controllable lateral size as reaction product.[7, 10-11] For example, ultrathin (1-2 nm) PbS NPLs were synthesized using the single-source precursor lead thiocyanate, and yielded NPLs with an orthorhombic crystal structure and high photoconductivity[11], while the absorption of PbS NPLs synthesized from lead octadecylxanthate was optically tuned, albeit showing weak NIR PL.[12]

In this work, we elucidate the characteristics and dynamics of photoexcited electronic states (excitons) in ultrathin PbS NPLs by using ultrafast transient optical absorption spectroscopy. We study cubic rock-salt phase PbS NPLs obtained by a near room temperature synthesis method using lead oleate and thiourea precursors (see **Scheme 1, Figure 1** and **S1** in the Supporting Information).

Synthesis conditions are chosen following Hendricks *et al.*[13] such that pure lead oleate is used, instead of producing it *in-situ* from oleic acid and lead acetate.[7, 14] The reaction is carried out in octylamine with oleic acid and thiourea as sulfur precursor. Exploiting an acid-base equilibrium between octylamine and oleic acid, which leads to the formation of an ammonium ($RNH_3^+$) species that supports the growth of 2D NPLs,[15] we obtain slightly elongated 2D PbS NPLs with a width of 7 nm and a length of 10 nm (see **Figure 1A**, **S1** and **Scheme 1**). The NPLs exhibit an absorption edge at 680 nm (1.8 eV) and NIR PL at 720 nm (1.7 eV) with a QY of 1.4 % (see **Figure 1B**).

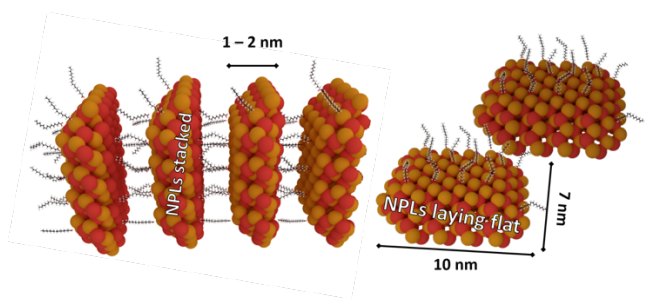

**Scheme 1.** Schematic of PbS NPLs stacked on edge and lying flat as observed in the TEM images shown in **Figure 1A**.

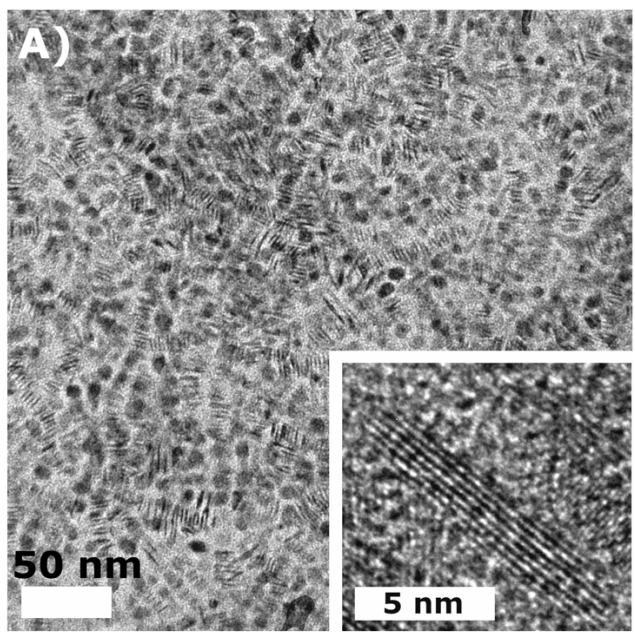

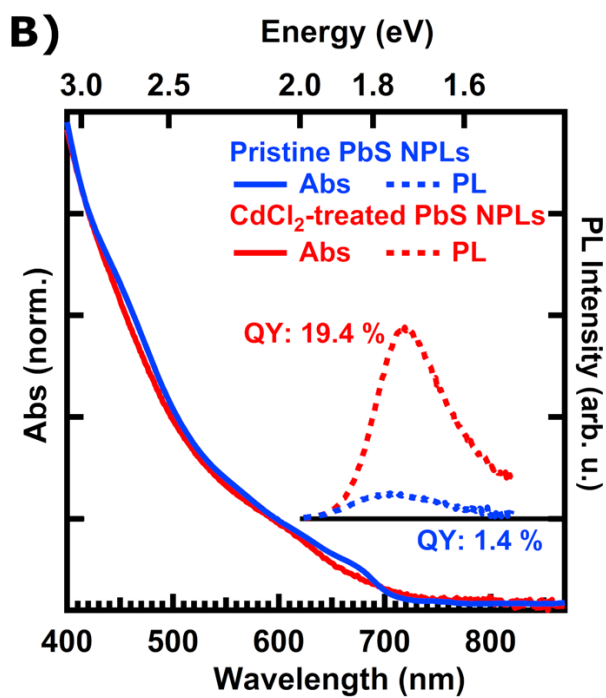

**Figure 1.** (A) TEM images of ultrathin PbS NPLs with lateral dimensions of 7 × 10 nm² and a thickness of 1-2 nm. Some NPLs are stacking on edge as shown in Scheme 1. The inset shows a single NPL on its edge. (B) Steady-state absorption and PL spectra of pristine PbS NPLs (blue) and after treatment with $CdCl_2$ (red), showing the absorption onset of the pristine NPLs at 680 nm (1.8 eV) and PL centered at 720 nm and broad absorption of the $CdCl_2$ treated sample. The photoluminescence quantum yield of $CdCl_2$ treated PbS NPLs reaches 19.4 %.

The obtained NPLs exhibit the typical reflexes of the cubic PbS crystal system (see **Figure S2** and **S3** for powder XRD and for SAED transmission electron microscopy rotational scans of the samples). EDS analysis shows a stoichiometric ratio of 1.1:1 for Pb:S in pristine NPL samples. In a post-synthetic step, the photoluminescence quantum yield (PLQY) of pristine PbS NPLs is increased to 19.4 % by treating the crude reaction product with a $CdCl_2$ solution in octylamine (see Supporting Info). Recently, Christodoulou *et al.* reported an increase in the thickness of CdSe NPLs after treating the samples with $CdCl_2$ with which the authors obtained CdSe NPLs with a thickness inaccessible under typical reaction conditions.[16] Generally, the surface treatment of NCs with metal halides leads to their stabilization and increase of their size.[17-21] In our case, after treating the pristine NPLs with $CdCl_2$, we find PbS layers with a slight regularity loss (see **Figure S4**) and an increased Pb:S ratio of 1.8:1 following EDS analysis, which resembles a lead rich surface typically found for PbS NCs.[22] A higher Pb:S ratio in $CdCl_2$ treated samples suggests the addition of Pb to the surface and/or edges of the NPLs, which is a reconstruction effect, as reported for CdSe NPLs[23] and NCs.[24] The similarity of the EDS spectra of pristine and $CdCl_2$ treated samples (see **Figure S5**) further excludes cation exchange which is e.g. observed in PbS NC samples.[25] **Figure 2** shows the steady-state PL spectra of pristine and $CdCl_2$ treated PbS NPLs in more detail. Generally, PbS NCs exhibit a rather broad PL at room temperature, which was attributed to two emissive states in the NCs by Caram *et al.*[26] The authors ascribe PL from the PbS NC band edge as well as PL from an emissive "pinned" trap state in NCs to be responsible for PL linewidth broadening. Indeed, we find PbS NPL PL spectra are best fitted with two Gaussians (see **Figure 2**, fit function in the Supporting Information) for pristine and $CdCl_2$ treated samples.

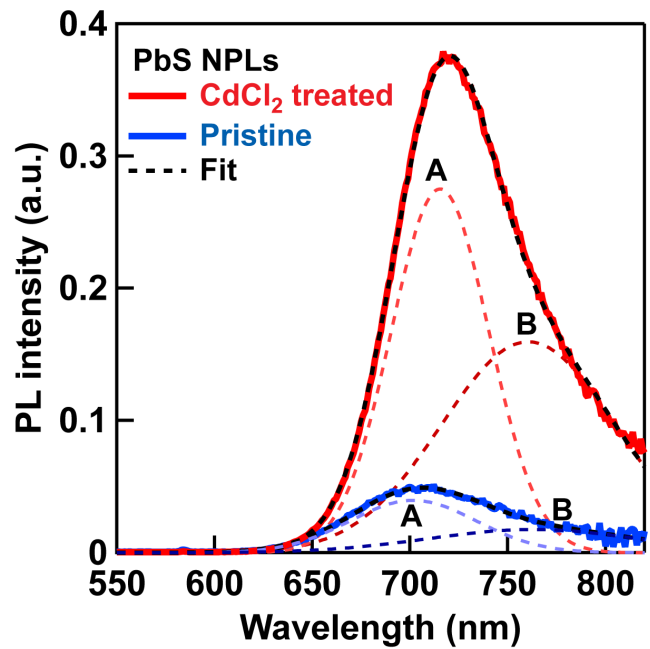

**Figure 2.** PL spectra of pristine (blue trace) and CdCl$_2$ treated (red trace) PbS NPLs. The emission is fit with a sum of two Gaussians (black dashed traces) with the two corresponding components A and B shown in lighter dashed lines.

**Table 1**. Values for the two Gaussian fits of the PbS NPL PL.

| PbS NPLs | Peak Position (nm) | | FWHM (nm) | | Peak area | | ratio |
|---|---|---|---|---|---|---|---|
| | peak A | peak B | peak A | peak B | peak A | peak B | A:B |
| Pristine | 701 | 760 | 75.5 | 127 | 3.13 | 2.47 | 1.27 |
| CdCl$_2$ treated | 715 | 760 | 59.1 | 105 | 17.3 | 17.7 | 0.98 |

Prior to CdCl$_2$ treatment, PbS NPLs exhibit a band gap associated PL peak labelled A at 701 nm with a FWHM of 75.5 nm (see **Table 1**). After the treatment, PL peak A at 715 nm shows a decreased FWHM of 59.1 nm. The "pinned" trap state associated peak labelled B stays fixed prior to and after the CdCl$_2$ treatment at 760 nm. Both PL contributions A and B are strongly increased after the CdCl$_2$ treatment, rendering it effective in increasing the overall PLQY of PbS NPLs.

As shown in **Figure 1,** CdCl$_2$ treated PbS NPLs exhibit a smaller absorption strength at 680 nm (1.8 eV) than pristine PbS NPLs and a similar absorption near 620 nm (2.0 eV).

To further characterize the nature and dynamics of photoexcited states in pristine and CdCl$_2$ treated PbS NPLs, we apply ultrafast transient optical absorption (TA) spectroscopy as described in refs.[27-30] and briefly in the Supporting Information. The samples are photoexcited close to the band edge at 700 nm to prevent dynamic effects by charge carrier cooling and with the same photoexcitation density ($2.5 \times 10^{13}$ photons/cm$^2$) so that the distribution of the number of excitons in both NPL samples is similar.

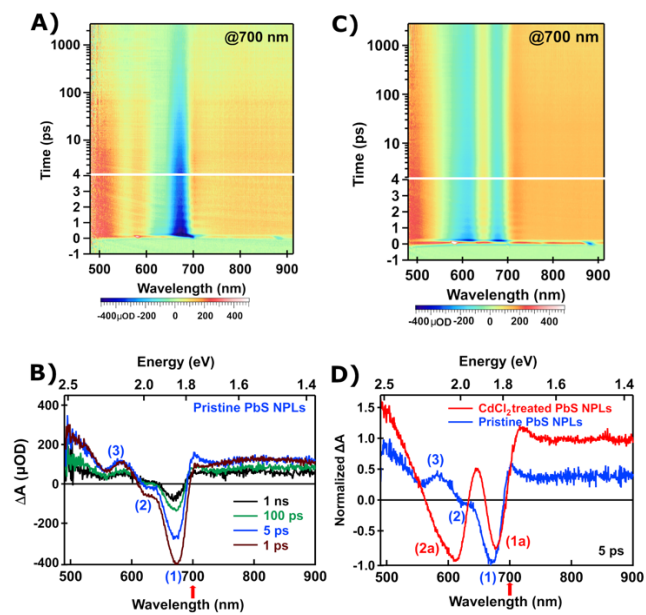

**Figure 3.** Color-coded 2D TA spectra of PbS NPLs photoexcited at 700 nm. (A) pristine PbS NPLs exhibit one prominent bleach feature (1) at 673 nm, features (2), (3) and spectral linecuts at different times after photoexcitation are plotted in (B) and discussed in the text. (C) CdCl$_2$ treated NPLs show a strong effect of the treatment, (D) includes spectral linecuts taken from (A) and (C) at a delay time of 5 ps after photoexcitation with CdCl$_2$ treated samples showing two prominent bleach features at 677 (1a) and 613 nm (2a), TA spectra are very distinct due to the involvement of different electronic states in the two samples as discussed in the text.

The shape of the TA spectra in **Figure 3** is discussed below on the basis of the terminology of electronic transitions and exciton states obtained from TA studies on PbSe[29, 31] and PbS NCs[28, 32].

**Figure 3A/C** shows 2D color-coded contour TA spectra of pristine and CdCl$_2$ treated PbS NPLs for comparison. The TA spectrum of pristine PbS NPLs exhibits a prominent negative ground-state (gs) bleach feature (1) of the 1S$_h$-1S$_e$ transition at 673 nm due to state filling and stimulated emission, also observed in the spectral linecuts in **Figure 3B**. The maximum of the gs bleach (1) is slightly blue-shifted with respect to the steady-state absorption at 680 nm (see **Figure 1**) implying a redshift in the excited state absorption due to the formation of biexcitons by the probe pulse.[31] The small bleach feature (2) originates from the formally forbidden 1S$_h$-1P$_e$ / 1P$_h$-1S$_e$ transitions that become slightly allowed in NCs due to symmetry breaking.[31, 33] At shorter wavelengths, we probe a derivative-like feature (3) that we attribute to a red-shift of the 1P$_h$-1P$_e$ transition superimposed on a red-shifted ground state absorption due to biexciton interactions.[34-36] We probe a wavelength-independent broad induced absorption at sub-band gap energies persisting over the TA time range studied (2.7 ns). This can be due to photoexcitation of trapped charges and/or due to further excitation of an exciton by the probe pulse, leading to intraband transitions.[37-39] The TA spectra of CdCl$_2$ treated PbS NPLs are very different from those of the pristine NPLs. Strikingly, they show **two** bleach features (1a and 2a) at 677 nm and 613 nm with similar amplitude and a long lifetime exceeding 3 ns (see **Figure 3C and 3D**). Since both bleach features are present for photoexcitation at a substantially longer wavelength (700 nm) than bleach feature (2a) at 613 nm, we exclude the latter being caused due to band edge excitons of thinner PbS NPLs after the CdCl$_2$ treatment. The strong bleach feature (2a) must be due to an additional electronic state introduced by the CdCl$_2$ treatment. In addition, the CdCl$_2$ treated PbS NPLs exhibit an increased induced absorption between features (1a) and (2a) and stronger sub-band gap trap state and intraband absorption.[37]

**Figure 4** includes the decay dynamics of the TA signal probed at the highest amplitude of the different bleach features for pristine and $CdCl_2$ treated PbS NPLs observed in **Figure 3**. The normalized decays show that bleach feature (1a) and (2a) level off at approximately 20 % of the original signal intensity after 20 ps, while bleach feature (1) decays over the time range of the measurement (15 % of the original signal intensity after 20 ps, resp. 5 % after 2.7 ns). The larger bleach at long times in $CdCl_2$ treated PbS NPLs agrees well with the found higher QY in the samples.

Interestingly, we probe coherent oscillations in the TA spectra of PbS NPLs at short times after photoexcitation as shown in the inset of **Figure 4**. Fourier transformation of the short-time decay curve residuals yields frequencies of 1.8 THz (7.4 meV, 60 cm$^{-1}$) for pristine and 2.2 THz (9.1 meV, 73 cm$^{-1}$) for $CdCl_2$ treated NPLs (see **Figure S6** and **S7**). At least for pristine PbS NPLs, a coupling of excitons to bulk-like phonon modes of PbS, which are known to occur at 1.96 THz (8.1 meV, 66 cm$^{-1}$ for the optical TO phonon)[40-41] cannot be completely excluded. However, the measured frequencies for the NPLs agree well with ranges reported earlier for strong exciton-phonon interactions of surface atoms in PbS NCs terminated by different ligands.[42-44] With ultrathin PbS NPLs exhibiting a high surface to thickness ratio with a considerable amount of surface atoms, we expect these interactions to dominate.

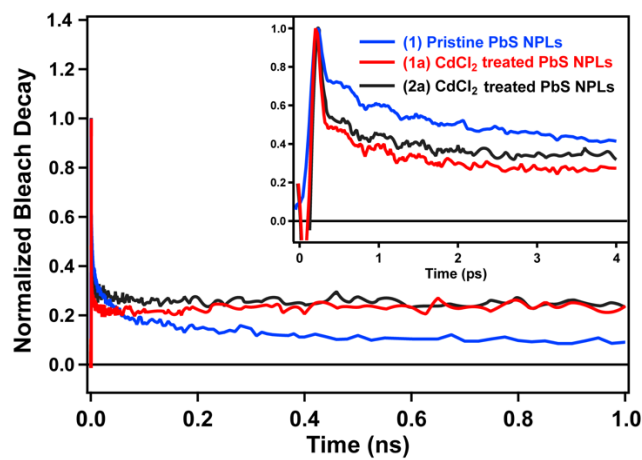

**Figure 4.** Decay dynamics of bleach features (1) of pristine and (1a, red), (2a, black) of $CdCl_2$ treated PbS NPLs with transients for pristine NPLs decaying over the course of the measurement, while $CdCl_2$ treated sample transients level off at 20 % of the original signal strength. The inset shows coherent oscillations originating from exciton-phonon interactions in PbS NPLs.

In conclusion, we have studied the optical properties of strongly confined colloidal cubic 2D PbS NPLs synthesized at near room temperature by steady-state PL and ultrafast TA spectroscopy for the first time. A new synthesis as well as a post-synthetic treatment of the obtained PbS NPLs with $CdCl_2$ increases their lead to sulfur ratio and leads to a strongly enhanced PLQY of 19.4 %. Additionally, coherent oscillations probed by TA originate from strong exciton-phonon interactions in PbS NPLs. Our results render the surface treatment as a valuable tool for tuning the optical properties of PbS NPLs for near-infrared emission.

ASSOCIATED CONTENT

**Supporting Information**.

The Supporting Information includes the synthesis of PbS NPLs, additional TEM images, XRD, EDS and PLQY analysis and brief description of the TA spectroscopic measurements.

AUTHOR INFORMATION

Corresponding Author

Dr. Jannika Lauth*, jannika.lauth@pci.uni-hannover.de

Author Contributions

The manuscript was written through contributions of all authors. All authors have given approval to the final version of the manuscript.

Funding Sources

L.F.K. and J.L. greatly acknowledge funding by the Deutsche Forschungsgemeinschaft (DFG, German Research Foundation) under Germany's Excellence Strategy within the Cluster of Excellence PhoenixD (EXC 2122, Project ID 390833453). J.L. is thankful for funding by the Caroline Herschel program of the Leibniz Universität Hannover.

ACKNOWLEDGMENT

We thank the Core Facility "Electron Microscopy" at the University of Oldenburg for access to a HR-TEM JEOL JEM FS2100 instrument. PD Dr. Erhard Rhiel and Prof. Dr. Katharina Al-Shamery are acknowledged for continuous support. Prof. Dr. Jürgen Caro and Prof. Dr. Armin Feldhoff are greatly acknowledged for the opportunity to perform XRD in their group.

REFERENCES

(1) Lauth, J.; Failla, M.; Klein, E.; Klinke, C.; Kinge, S.; Siebbeles, L. D. A. Photoexcitation of PbS Nanosheets Leads to Highly Mobile Charge Carriers and Stable Excitons. *Nanoscale* **2019**, *11*, 21569-21576.
(2) Aerts, M.; Bielewicz, T.; Klinke, C.; Grozema, F. C.; Houtepen, A. J.; Schins, J. M.; Siebbeles, L. D. A. Highly Efficient Carrier Multiplication in PbS Nanosheets. *Nat. Commun.* **2014**, *5*, 3789.
(3) Steiner, A. M.; Lissel, F.; Fery, A.; Lauth, J.; Scheele, M. Prospects of Coupled Organic-Inorganic Nanostructures for Charge and Energy Transfer Applications. *Angew. Chem. Int. Ed.* **2020**, DOI: **10.1002/anie.201916402**
(4) Schliehe, C.; Juarez, B. H.; Pelletier, M.; Jander, S.; Greshnykh, D.; Nagel, M.; Meyer, A.; Foerster, S.; Kornowski, A.; Klinke, C.; Weller, H. Ultrathin PbS Sheets by Two-Dimensional Oriented Attachment. *Science* **2010**, *329*, 550-553.
(5) Bielewicz, T.; Ramin Moayed, M. M.; Lebedeva, V.; Strelow, C.; Rieckmann, A.; Klinke, C. From Dots to Stripes to Sheets: Shape Control of Lead Sulfide Nanostructures. *Chem. Mater.* **2015**, *27*, 8248-8254.
(6) Bielewicz, T.; Dogan, S.; Klinke, C. Tailoring the Height of Ultrathin PbS Nanosheets and Their Application as Field-Effect Transistors. *Small* **2015**, *11*, 826-833.
(7) Morrison, P. J.; Loomis, R. A.; Buhro, W. E. Synthesis and Growth Mechanism of Lead Sulfide Quantum Platelets in Lamellar Mesophase Templates. *Chem. Mater.* **2014**, *26*, 5012-5019.
(8) Riedinger, A.; Ott, F. D.; Mule, A.; Mazzotti, S.; Knusel, P. N.; Kress, S. J. P.; Prins, F.; Erwin, S. C.; Norris, D. J. An Intrinsic Growth Instability in Isotropic Materials leads to Quasi-Two-Dimensional Nanoplatelets. *Nat. Mater.* **2017,** *16*, 743.
(9) Ott, F. D.; Riedinger, A.; Ochsenbein, D. R.; Knüsel, P. N.; Erwin, S. C.; Mazzotti, M.; Norris, D. J. Ripening of Semiconductor Nanoplatelets. *Nano Lett.* **2017**, *17*, 6870-6877.
(10) Khan, A. H.; Pal, S.; Dalui, A.; Pradhan, J.; Sarma, D. D.; Acharya, S. Solution-Processed Free-Standing Ultrathin Two-Dimensional PbS Nanocrystals with Efficient and Highly Stable Dielectric Properties. *Chem. Mater.* **2017,** *29*, 1175-1182.
(11) Akkerman, Q. A.; Martín-García, B.; Buha, J.; Almeida, G.; Toso, S.; Marras, S.; Bonaccorso, F.; Petralanda, U.; Infante, I.; Manna, L. Ultrathin Orthorhombic PbS Nanosheets. *Chem. Mater.* **2019,** *31*, 8145-8153.
(12) Khan, A. H.; Brescia, R.; Polovitsyn, A.; Angeloni, I.; Martín-García, B.; Moreels, I. Near-Infrared Emitting Colloidal PbS Nanoplatelets: Lateral Size Control and Optical Spectroscopy. *Chem. Mater.* **2017,** *29*, 2883-2889.
(13) Hendricks, M. P.; Campos, M. P.; Cleveland, G. T.; Jen-La Plante, I.; Owen, J. S. A Tunable Library of Substituted Thiourea Precursors to Metal Sulfide Nanocrystals. *Science* **2015**, *348*, 1226-1230.
(14) Houtepen, A. J.; Koole, R.; Vanmaekelbergh, D.; Meeldijk, J.; Hickey, S. G. The Hidden Role of Acetate in the PbSe Nanocrystal Synthesis. *J. Am. Chem. Soc.* **2006**, *128*, 6792-6793.
(15) Almeida, G.; Goldoni, L.; Akkerman, Q.; Dang, Z.; Khan, A. H.; Marras, S.; Moreels, I.; Manna, L. Role of Acid–Base Equilibria in the Size, Shape, and Phase Control of Cesium Lead Bromide Nanocrystals. *ACS Nano* **2018**, *12*, 1704-1711.
(16) Christodoulou, S.; Climente, J. I.; Planelles, J.; Brescia, R.; Prato, M.; Martin-Garcia, B.; Khan, A. H.; Moreels, I. Chloride-Induced Thickness Control in CdSe Nanoplatelets. *Nano Lett* **2018,** *18*, 6248-6254.

(17) Zhang, J.; Gao, J.; Miller, E. M.; Luther, J. M.; Beard, M. C. Diffusion-Controlled Synthesis of PbS and PbSe Quantum Dots with in Situ Halide Passivation for Quantum Dot Solar Cells. *ACS Nano* **2014,** *8*, 614-622.
(18) Lim, J.; Bae, W. K.; Park, K. U.; zur Borg, L.; Zentel, R.; Lee, S.; Char, K. Controlled Synthesis of CdSe Tetrapods with High Morphological Uniformity by the Persistent Kinetic Growth and the Halide-Mediated Phase Transformation. *Chem. Mater.* **2013,** *25*, 1443-1449.
(19) Crisp, R. W.; Kroupa, D. M.; Marshall, A. R.; Miller, E. M.; Zhang, J.; Beard, M. C.; Luther, J. M. Metal Halide Solid-State Surface Treatment for High Efficiency PbS and PbSe QD Solar Cells. *Sci. Rep.* **2015,** *5*, 9945.
(20) Cho, W.; Kim, S.; Coropceanu, I.; Srivastava, V.; Diroll, B. T.; Hazarika, A.; Fedin, I.; Galli, G.; Schaller, R. D.; Talapin, D. V. Direct Synthesis of Six-Monolayer (1.9 nm) Thick Zinc-Blende CdSe Nanoplatelets Emitting at 585 nm. *Chem. Mater.* **2018,** *30*, 6957-6960.
(21) Meerbach, C.; Wu, C.; Erwin, S. C.; Dang, Z.; Prudnikau, A.; Lesnyak, V. Halide-Assisted Synthesis of Cadmium Chalcogenide Nanoplatelets. *Chem. Mater.* **2020,** *32*, 566-574.
(22) Weidman, M. C.; Beck, M. E.; Hoffman, R. S.; Prins, F.; Tisdale, W. A. Monodisperse, Air-Stable PbS Nanocrystals via Precursor Stoichiometry Control. *ACS Nano* **2014,** *8*, 6363-6371.
(23) Sun, H.; Buhro, W. E. Reversible Z-Type to L-Type Ligand Exchange on Zinc-Blende Cadmium Selenide Nanoplatelets. *Chem. Mater.* **2020**, *32*, 5814-5826.
(24) Marino, E.; Kodger, T. E.; Crisp, R. W.; Timmerman, D.; MacArthur, K. E.; Heggen, M.; Schall, P. Repairing Nanoparticle Surface Defects. *Angew. Chem. Int. Ed.* **2017,** *56*, 13795-13799.
(25) Zhang, J.; Chernomordik, B. D.; Crisp, R. W.; Kroupa, D. M.; Luther, J. M.; Miller, E. M.; Gao, J.; Beard, M. C. Preparation of Cd/Pb Chalcogenide Heterostructured Janus Particles via Controllable Cation Exchange. *ACS Nano* **2015,** *9*, 7151-7163.
(26) Caram, J. R.; Bertram, S. N.; Utzat, H.; Hess, W. R.; Carr, J. A.; Bischof, T. S.; Beyler, A. P.; Wilson, M. W. B.; Bawendi, M. G. PbS Nanocrystal Emission Is Governed by Multiple Emissive States. *Nano Lett.* **2016,** *16*, 6070-6077.
(27) Lauth, J.; Kinge, S.; Siebbeles, L. D. A. Ultrafast Transient Absorption and Terahertz Spectroscopy as Tools to Probe Photoexcited States and Dynamics in Colloidal 2D Nanostructures. *Z. Phys. Chem.* **2017,** *231*, 107-119.
(28) Lauth, J.; Grimaldi, G.; Kinge, S.; Houtepen, A. J.; Siebbeles, L. D. A.; Scheele, M. Ultrafast Charge Transfer and Upconversion in Zn β-Tetraaminophthalocyanine Functionalized PbS Nanostructures Probed by Transient Absorption Spectroscopy. *Angew. Chem. Int. Ed.* **2017,** *56*, 14061-14065.
(29) Spoor, F. C. M.; Kunneman, L. T.; Evers, W. H.; Renaud, N.; Grozema, F. C.; Houtepen, A. J.; Siebbeles, L. D. A. Hole Cooling Is Much Faster than Electron Cooling in PbSe Quantum Dots. *ACS Nano* **2016**, *10*, 695-703.
(30) Kunneman, L. T.; Tessier, M. D.; Heuclin, H.; Dubertret, B.; Aulin, Y. V.; Grozema, F. C.; Schins, J. M.; Siebbeles, L. D. A. Bimolecular Auger Recombination of Electron–Hole Pairs in Two-Dimensional CdSe and CdSe/CdZnS Core/Shell Nanoplatelets. *J. Phys. Chem. Lett.* **2013,** *4*, 3574-3578.
(31) Schins, J. M.; Trinh, M. T.; Houtepen, A. J.; Siebbeles, L. D. A. Probing Formally Forbidden Optical Transitions in PbSe Nanocrystals by Time- and Energy-Resolved Transient Absorption Spectroscopy. *Phys. Rev. B* **2009,** *80*, 035323.
(32) Yang, Y.; Rodríguez-Córdoba, W.; Lian, T. Ultrafast Charge Separation and Recombination Dynamics in Lead Sulfide Quantum Dot–Methylene Blue Complexes Probed by Electron and Hole Intraband Transitions. *J. Am. Chem. Soc.* **2011,** *133*, 9246-9249.

(33)    Trinh, M. T.; Houtepen, A. J.; Schins, J. M.; Piris, J.; Siebbeles, L. D. A. Nature of the Second Optical Transition in PbSe Nanocrystals. *Nano Lett*. **2008,** *8*, 2112-2117.
(34)    Geiregat, P.; Houtepen, A.; Justo, Y.; Grozema, F. C.; Van Thourhout, D.; Hens, Z. Coulomb Shifts upon Exciton Addition to Photoexcited PbS Colloidal Quantum Dots. *J. Phys. Chem. C* **2014,** *118*, 22284-22290.
(35)    Gdor, I.; Shapiro, A.; Yang, C.; Yanover, D.; Lifshitz, E.; Ruhman, S. Three-Pulse Femtosecond Spectroscopy of PbSe Nanocrystals: 1S Bleach Nonlinearity and Sub-Band-Edge Excited-State Absorption Assignment. *ACS Nano* **2015,** *9*, 2138-2147.
(36)    Kambhampati, P. Unraveling the Structure and Dynamics of Excitons in Semiconductor Quantum Dots. *Acc. Chem. Res*. **2011,** *44*, 1-13.
(37)    De Geyter, B.; Houtepen, A. J.; Carrillo, S.; Geiregat, P.; Gao, Y.; ten Cate, S.; Schins, J. M.; Van Thourhout, D.; Delerue, C.; Siebbeles, L. D. A.; Hens, Z. Broadband and Picosecond Intraband Absorption in Lead-Based Colloidal Quantum Dots. *ACS Nano* **2012,** *6*, 6067-6074.
(38)    Geiregat, P.; Houtepen, A. J.; Van Thourhout, D.; Hens, Z. All-Optical Wavelength Conversion by Picosecond Burst Absorption in Colloidal PbS Quantum Dots. *ACS Nano* **2016,** *10*, 1265-1272.
(39)    An, J. M.; Franceschetti, A.; Dudiy, S. V.; Zunger, A. The Peculiar Electronic Structure of PbSe Quantum Dots. *Nano Lett*. **2006,** *6*, 2728-2735.
(40)    Elcombe, M. M. The Crystal Dynamics of Lead Sulphide *Proc. R. Soc. London Ser. A* **1967,** *300*, 210-217.
(41)    Machol, J. L.; Wise, F. W.; Patel, R. C.; Tanner, D. B. Vibronic Quantum Beats in PbS Microcrystallites. *Phys. Rev. B* **1993,** *48*, 2819-2822.
(42)    Seydel, T.; Koza, M. M.; Matsarskaia, O.; André, A.; Maiti, S.; Weber, M.; Schweins, R.; Prévost, S.; Schreiber, F.; Scheele, M. A Neutron Scattering Perspective on the Structure, Softness and Dynamics of the Ligand Shell of PbS Nanocrystals in Solution. *Chem. Sci*. **2020,** *11*, 8875-8884.
(43)    Yazdani, N.; Bozyigit, D.; Vuttivorakulchai, K.; Luisier, M.; Infante, I.; Wood, V. Tuning Electron–Phonon Interactions in Nanocrystals through Surface Termination. *Nano Lett*. **2018**, *18*, 2233-2242.
44.    Bozyigit, D.; Yazdani, N.; Yarema, M.; Yarema, O.; Lin, W. M. M.; Volk, S.; Vuttivorakulchai, K.; Luisier, M.; Juranyi, F.; Wood, V. Soft Surfaces of Nanomaterials Enable Strong Phonon Interactions. *Nature* **2016,** *531*, 618-622.

# Supporting Information

# Probing Excitons in Ultrathin PbS Nanoplatelets with Enhanced Near-Infrared Emission

*Francisco Manteiga Vázquez,[†] Qianli Yu,[†] Lars F. Klepzig,[+//] Laurens D. A. Siebbeles,[†] Ryan W. Crisp,[†‡] Jannika Lauth [+//†*]*

[†]Chemical Engineering Department, Delft University of Technology, Van der Maasweg 9, NL-2629 HZ Delft, The Netherlands

[+]Institute of Physical Chemistry and Electrochemistry, Leibniz Universität Hannover, Callinstr. 3A, D-30167 Hannover, Germany

[//]Cluster of Excellence PhoenixD (Photonics, Optics, and Engineering – Innovation Across Disciplines), Hannover, Germany

[‡]Chemistry of Thin Film Materials, Department of Chemistry and Pharmacy, Friedrich-Alexander University Erlangen-Nürnberg, Cauerstr. 3, D-91058, Erlangen, Germany

Corresponding Author

*Dr. Jannika Lauth, jannika.lauth@pci.uni-hannover.de

Synthesis of cubic ultrathin PbS NPLs

*Sulfur precursor*

360 mg (4.7 mmol) of thiourea was weighed into a three-neck flask **(1)** under ambient conditions. The flask was equipped with a glass insert for a thermocouple, a septum and then connected to a Schlenk-apparatus and set under nitrogen flow. 9 mL (54.5 mmol) of *n*-octylamine was injected into the flask to dissolve thiourea with the temperature set at 35 °C. The solution was stirred at 500 rpm for at least 18 hours before using.

*Lead precursor*

Pb(oleate)$_2$ was synthesized by a method described by Hendricks *et. al.*[S1] For the PbS NPL synthesis, 365 mg (0.475 mmol) Pb(oleate)$_2$ was weighed into a three-neck flask **(2)** under inert gas (N$_2$) and subsequently kept at the Schlenk-apparatus under nitrogen atmosphere. 3 mL (18.2 mmol) of *n*-octylamine and 1.6 mL (5 mmol) oleic acid were added to the flask.

*PbS NPL Synthesis*

At ~35 °C, the PbS NPL reaction is started in flask **(2)** by adding 0.5 mL (0.3 mmol) of the thiourea precursor from flask **(1)**. The reaction is run for 16-18 minutes at a temperature range between 32 – 38 °C. The color of the solution changes from yellow to light orange instantly when thiourea is injected into the lead precursor. The color gradually darkens to dark orange/brown.

PbS NPLs are precipitated once by injecting the crude product under inert gas conditions into a vial containing a 1:2 mixture of dry toluene:ethanol and centrifugation at 5000 rpm for 5 minutes. Hexane is used for redispersing the NPLs.

CdCl$_2$ Treatment of PbS NPLs

367 mg (2 mmol) CdCl$_2$ are weighed into a three-neck flask, which is then connected to the Schlenk-apparatus. 20 mL of *n*-octylamine are subsequently added and the turbid solution set stirring at 80 °C under mild vacuum (~0.1 mbar) for 30 min. After setting the CdCl$_2$ solution under nitrogen atmosphere, the temperature is raised to 120 °C for another 30 minutes and then allowed to cool to 35 °C. This solution is kept stirring overnight, before being used.

Typically, 5 mL (0.5 mmol) of this CdCl$_2$ solution in octylamine are added after running the PbS NPL synthesis for 16 min at 35 °C. The reaction is kept stirring for 5-8 more minutes before it is quenched by a water bath and precipitated like described before.

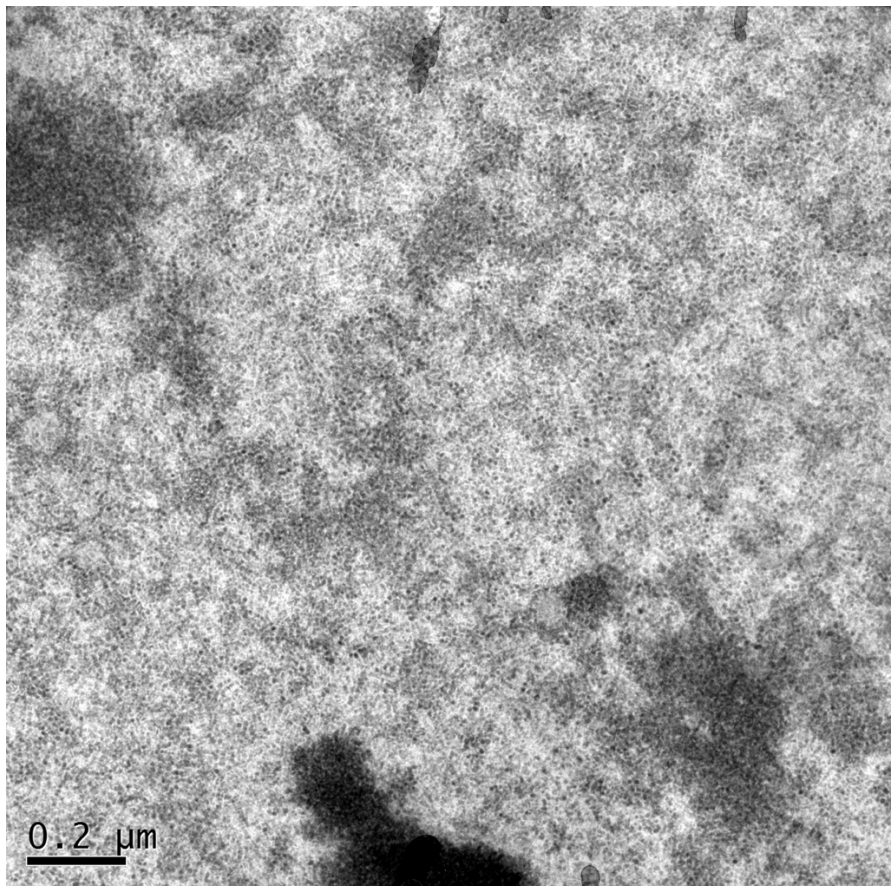

**Figure S1.** TEM overview on a dense PbS NPL film deposited from solution.

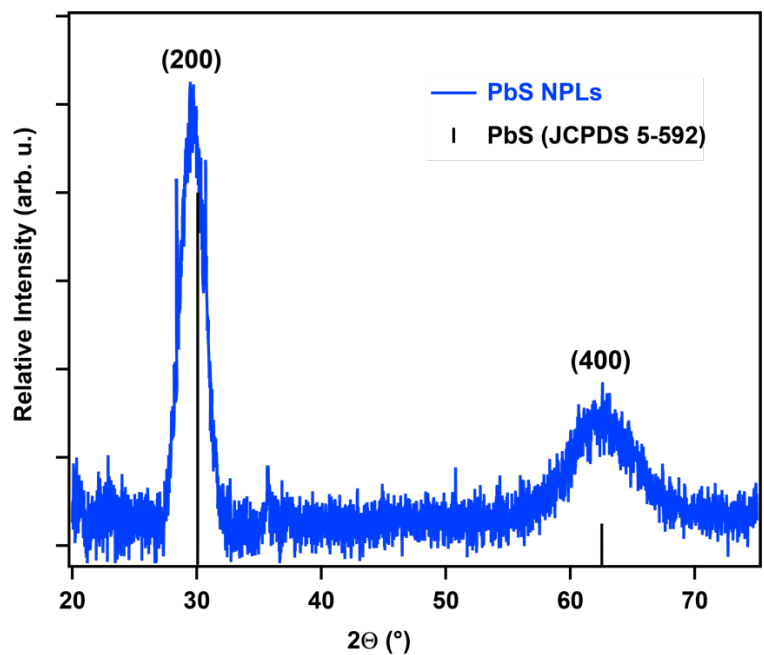

**Figure S2.** XRD of PbS NPLs exhibiting the cubic rock-salt structure of the materials as has been described by Bielewicz *et al*.[S2]

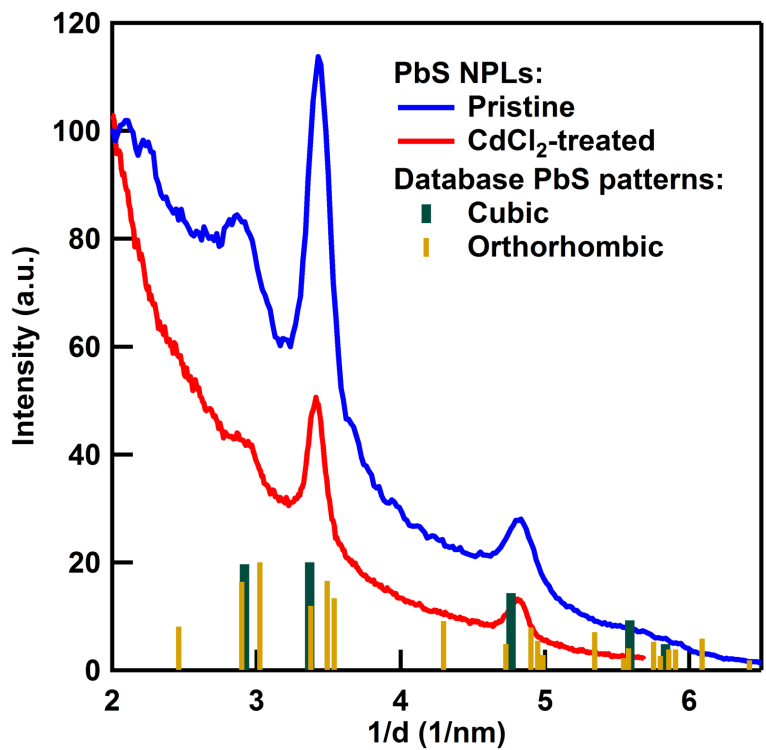

**Figure S3.** Rotational scan intensity profile of selected area electron diffraction (SAED) of an ensemble of PbS NPLs with first three reflections best fitted with the cubic crystal system, ICSD files: cubic ICSD38293, orthorhombic ICSD648451.

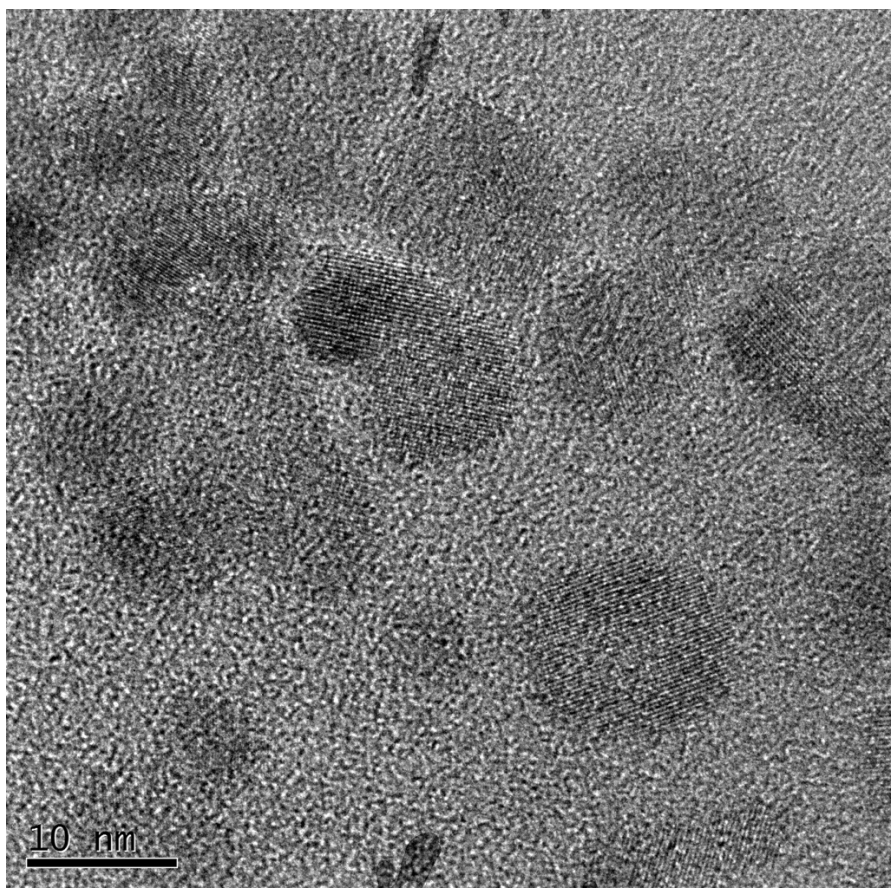

**Figure S4.** After CdCl$_2$ treatment PbS NPLs exhibit a slight loss of regularity and broadening of optical features as shown in the main manuscript.[S3]

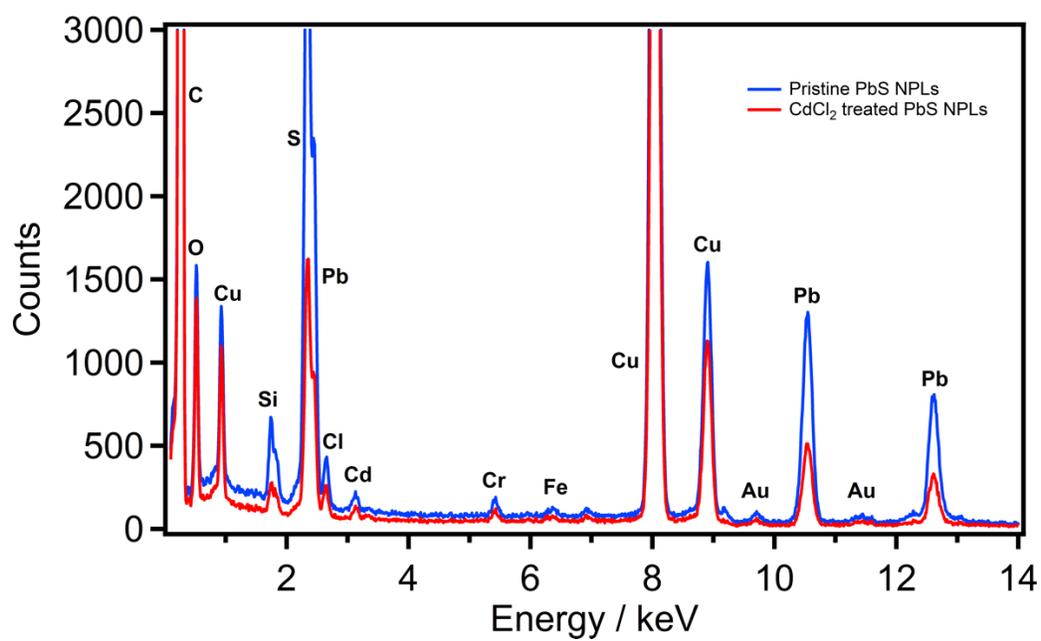

**Figure S5.** Apart from the lead to sulfur ratio, EDS of pristine PbS NPLs and CdCl$_2$ treated samples look similar so that cation exchange from lead to cadmium is excluded.[S3]

**Photoluminescence and quantum yield measurements**

The PL of pristine and CdCl$_2$ treated PbS NPLs was measured with a FLS980 PL spectrometer by Edinburgh Instruments. The samples were prepared by dilution in hexane in cuvettes with 10 mm path length. A time of 60 min was kept before the measurements to allow heating up of the spectrometer lamp and to minimize any fluctuations in the measuring conditions. Unprecipitated samples taken directly from the reaction solution were measured and excited at 500 nm.

$$f(x) = A e^{-\frac{(x-x_A)^2}{\sigma_A^2}} + B e^{-\frac{(x-x_B)^2}{\sigma_B^2}}$$

The evaluation of the photoluminescence quantum yield (PLQY) was performed using a relative method described by Würth et al.[S4] Rhodamin 6G (Rh6G) in dried ethanol is used as reference with a known quantum yield of 0.92. The PLQY is given by

$$PLQY = 0.92 \cdot \frac{A_{sample}}{A_{Rh6G}} \cdot \frac{1 - 10^{-Abs_{Rh6G}}}{1 - 10^{-Abs_{sample}}} \cdot \left(\frac{n_{Hex}}{n_{EtOH}}\right)^2$$

with *A* being the integral area of the PL measurement curve, *Abs* the absorbance, and *n* the refractive index of the solvent, each indexed with the sample or Rh6G for reference.

**Transient absorption spectroscopy**

PbS NPL were studied by broadband pump-probe ultrafast transient optical absorption (TA) spectroscopy in a set-up described previously and briefly discussed here.[S5] Solutions were analyzed in 2 mm quartz cuvettes and kept under constant stirring to prevent photocharging.[S6] For the set-up, 180 fs laser pulses from a Yb:KGW oscillator with a fundamental wavelength at 1028 nm are beam-split to generate a pump and a probe beam. The pump beam energy is varied by nonlinear frequency mixing in an optical parametric amplifier (OPA) and second harmonics generation (Light Conversion, Orpheus). A broadband probe spectrum is generated by focusing the 1028 nm laser light onto a sapphire (500-1500 nm) or a CaF$_2$ (400 – 600 nm) crystal by nonlinear processes.

The pump pulse is dumped after the photoexcitation of the sample, while the probe light is led to a detector fiber suitable for the probe spectrum selected (Helios, Ultrafast Systems).

**Exciton-phonon interactions in PbS NPLs**

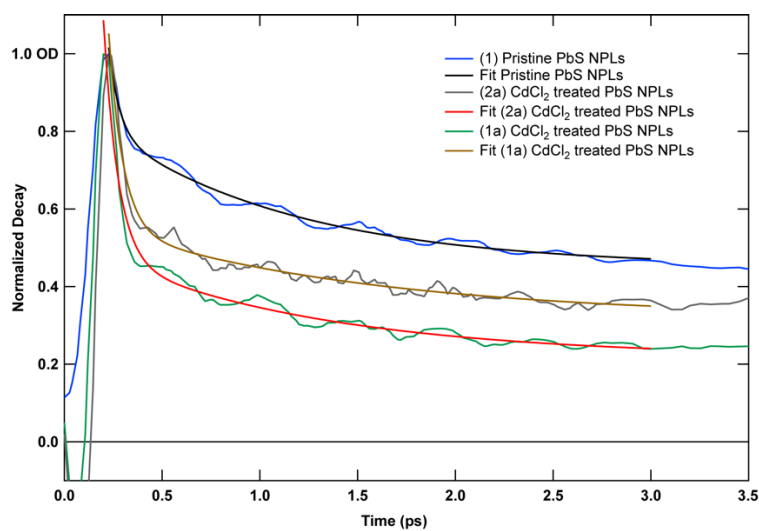

**Figure S6.** Biexponential fits that are subtracted from coherent oscillations of high PED TA measurements of pristine and CdCl$_2$ treated PbS NPLs.

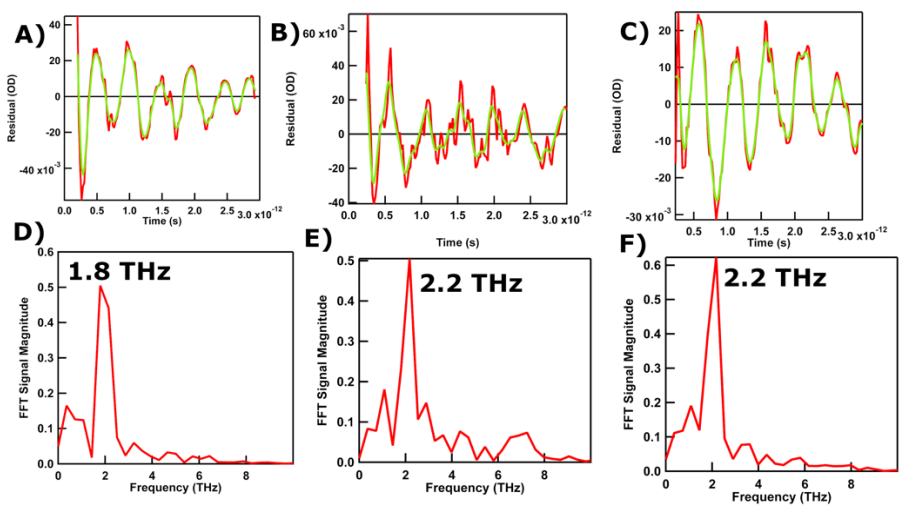

**Figure S7.** Fit residuals (binomially smoothed) of the decay curves shown in **Figure S6**, A-C) (1) pristine and (2a) and (1a) CdCl$_2$ treated PbS NPLs, D-F) associated FFTs with frequencies of the oscillation of 1.8 THz for pristine and 2.2 THz for CdCl$_2$ treated PbS NPLs.

References

(S1)   Hendricks, M. P.; Campos, M. P.; Cleveland, G. T.; Jen-La Plante, I.; Owen, J. S. A Tunable Library of Substituted Thiourea Precursors to Metal Sulfide Nanocrystals. *Science* **2015,** *348*, 1226-1230.
(S2)   Bielewicz, T.; Dogan, S.; Klinke, C. Tailoring the Height of Ultrathin PbS Nanosheets and Their Application as Field-Effect Transistors. *Small* **2015,** *11*, 826-833.
(S3)   Christodoulou, S.; Climente, J. I.; Planelles, J.; Brescia, R.; Prato, M.; Martín-García, B.; Khan, A. H.; Moreels, I. Chloride-Induced Thickness Control in CdSe Nanoplatelets. *Nano Lett*. **2018,** *18*, 6248-6254.
(S4)   Würth, C.; Grabolle, M.; Pauli, J.; Spieles, M.; Resch-Genger, U. Relative and Absolute Determination of Fluorescence Quantum Yields of Transparent Samples. *Nat*. *Protoc*. **2013,** *8*, 1535-1550.
(S5)   Lauth, J.; Grimaldi, G.; Kinge, S.; Houtepen, A. J.; Siebbeles, L. D. A.; Scheele, M. Ultrafast Charge Transfer and Upconversion in Zn β-Tetraaminophthalocyanine Functionalized PbS Nanostructures Probed by Transient Absorption Spectroscopy. *Angew*. *Chem*. *Int*. *Ed*. **2017,** *56*, 14061-14065.
(S6)   McGuire, J. A.; Joo, J.; Pietryga, J. M.; Schaller, R. D.; Klimov, V. I. New Aspects of Carrier Multiplication in Semiconductor Nanocrystals. *Acc*. *Chem*. *Res*. **2008,** *41*, 1810-1819.